# Gusev's stochastic model for the seismic source: high-frequency behavior in the far zone

G.M. Molchan

*Institute of Earthquake Prediction Theory and*
*Mathematical Geophysics, Russian Academy of Sciences,*
*Moscow*

**Abstract.** This paper discusses local features in the source function that generate in the far zone simultaneously (a) a quadratic decay of the source spectrum and (b) the loss of radiation directivity at high frequencies. A. Gusev drew attention to this problem and suggested that a positive solution can be obtained for earthquake rupture front with a rather complex ("lace") structure. Below we give a theoretical solution of the problem and show that the front structure can be simple enough, but not smooth.

.

1. **Introduction.**

A. Gusev [2013, 2014, 2018] drew attention to a problem of the modeling of the seismic source. The far-zone elastic displacement for this source must have two properties: (a) $\omega^{-2}$ spectrum decaying at high frequencies (HF) and (b) the independence of HF radiation of the direction to the receiver. The first of these properties will be referred to as the omega-square behavior and the second the loss of radiation directivity. The quadratic decay of the spectrum is a common assumption in engineering practice, although it is acceptable that its frequency range may be limited [Gusev et al., 2017]. A similar situation exist with respect to the universality of property (b). The corresponding counterexamples can be found in [Kurzon et al., 2014; Wen et al., 2014], related to the frequency range $\omega > \omega_0$, where $\omega_0$ is the Brune [1970] corner frequency. Any exceptions to the rule emphasize the complexity of the Gusev problem and call for an understanding of the conditions under which these properties become feasible.

Gusev proposed two kinematic source models, essentially very similar ones, involving some elements of stochastic behavior [Gusev, 2014, 2018]. These models have made it possible to reproduce the key properties of HF radiation: the existence of two corner frequencies, a plateau in acceleration source spectrum above the Brune's corner frequency (which is equivalent to property (a)), and a weak effect of radiation directivity (in the 2014 model). The fractality of the rupture front plays an important part in these models. To reproduce the key properties, the rupture front in the Gusev models is multiply connected ("lacy") fractal polyline that occupies a strip of finite width. Since the analysis and fitting of the models were done numerically, the issue of the origin of properties (a, b) remains an open question. It is, in particular, not clear just how complex must be the lace structure of a rupture front for (a, b) to occur.

Below we are going to show that property (a) can be generated by slightly non- smooth rupture front, while property (b) can be reproduced in conditions of smoothness of distributions of an irregular noise-like component of the rupture front. As a result, we can conclude that fractal models of the seismic source function unexpectedly complement classical smooth source models (Aki and Richards, 1980). More exactly, property (a) is unstable in the classical models, while property (b) is as a rule altogether absent. However, the addition of a slight irregular component to the front enables one to remove this instability and to also produce property (b). This observation is contained in a technically complicated paper (Molchan, 2015). The analysis to be presented below is more general, as well as being simpler.

The rest of this paper is organized as follows: Section 3 describes local features in a smooth source function that can be responsible for the quadratic decay of the spectrum. Section 4 generalizes the class of doubly stochastic Gusev models and contains theoretical conditions for properties (a, b). In section 5, the derived spectral asymptotics are discussed informally. Reading is facilitated by relegating the mathematical statements to the Appendix.

## 2. Preliminaries.

It is well known (Aki and Richards, 1980) that the *far-zone* displacement field can be represented as follows:

$$u(t) = A\int_\Sigma f(\mathbf{g}, t - dist(\mathbf{g}, G_{rec})/c)d\Sigma \approx A\int_\Sigma f(\mathbf{g}, t - t_0 + <\mathbf{g}, \gamma_\Sigma>/c))d\mathbf{g}. \qquad (1)$$

Here, $G_{rec}$ is the receiver location, $\Sigma$ is the rupture plane whose coordinates are $(g_1, g_2) = \mathbf{g}$, $f$ is the source function, i. e., local displacement velocity $\Delta \dot{u}(\mathbf{g}, t)$, $\gamma_\Sigma$ is the orthogonal projection of the hypocenter-receiver direction, $\gamma$, onto the plane $\Sigma$; $<\cdot,\cdot>$ denotes a scalar product, $c$ is wave velocity, $t_0 c = dist(hypocenter, G_{rec})$, and $d\mathbf{g} = dg_1 dg_2$ is an element of area. For simplicity, the representation (1) is scalar.

The Fourier transform can be applied to (1) to derive a dual spectral representation:

$$\hat{u}(\omega) \approx A\int d\Sigma \int e^{i\omega t} f(\mathbf{g}, t - t_0 + <\mathbf{g}, \gamma_\Sigma>/c)dt$$

$$= A e^{i\omega t_0} \int d\Sigma \int e^{i\omega(s - <\mathbf{g}, \gamma_\Sigma>/c)} f(\mathbf{g}, s) ds = A e^{i\omega t_0} \hat{f}(-\gamma_\Sigma c^{-1}\omega, \omega) \qquad (2)$$

Here, $\hat{f}(\mathbf{k}, \omega)$ is the space-time Fourier transform of function $f(\mathbf{x}) := f(\mathbf{g}, t)$ which is extended as zero outside its domain of definition (*support*) $\Omega_f$.

The problem of the omega-square behavior for the spectrum is stated as follows: describe a physically natural class of source functions $f(\mathbf{x})$ for which $\hat{f}(\mathbf{p}\omega)$, $\mathbf{p} = (-\gamma_\Sigma/c, 1)$ has a stable $\omega^{-2}$ asymptotics. The term 'stable' means that the asymptotic behavior takes place for a rather massive set of directions.

Since the Fourier transform is linear, it is simpler to study local fragments of $f(\mathbf{x})$, namely, $f(\mathbf{x}|\mathbf{x}_{0i}) = f(\mathbf{x})\varphi(\mathbf{x}|\mathbf{x}_{0i})$. Here, the $\varphi(\mathbf{x}|\mathbf{x}_{0i})$ are auxiliary infinitely

smooth functions: they are zero outside a small vicinity of $\mathbf{x}_{0i}$, $O_{x_{0i}}$, are 1 in a smaller vicinity $0.5 O_{x_{0i}}$, and their sum is one: $\sum \varphi(\mathbf{x}|\mathbf{x}_{0i}) = 1$. These fragments do not distort $f(\mathbf{x})$ near selected points $\mathbf{x}_{0i}$ and make it vanish outside the small vicinities in a smooth manner. For any fragment $f(\mathbf{x}|\mathbf{x}_0)$ we shall treat the HF asymptotics of the integral

$$\int e^{i\omega<\mathbf{p},\mathbf{x}>} f(\mathbf{x})\varphi(\mathbf{x}|\mathbf{x}_0) d\mathbf{x} := \hat{f}(\mathbf{p}\omega|\mathbf{x}_0) \qquad (3)$$

as the contribution of $\mathbf{x}_0$ into the asymptotics of $\hat{u}(\omega)$. The decomposition of a source function into fragments makes it possible to describe the spectral asymptotics in terms of possible local features observable in $f(\mathbf{x})$. In this way, we need not specify the model $f(\mathbf{x})$ as a whole. In addition, the features of $f(\mathbf{x})$ which are discussed below have point-like in nature and therefore their spectral contribution (3) will not depend on the choice of auxiliary functions $\varphi(\mathbf{x}|\mathbf{x}_{0i})$.

Below we consider two types of local features in source functions, piecewise smooth and fractal ones.

## 3. Smooth source functions.

In this section, the source function $f(\mathbf{x})$, $\mathbf{x} = (\mathbf{g},t)$ is assumed to be smooth within its bounded support $\Omega_f \in R^3$. Therefore for any inner point $\mathbf{x}_0 = (\mathbf{g}_0, t_0)$ of $\Omega_f$, the fragment $f(\mathbf{x}|\mathbf{x}_0)$ can be considered as a smooth function with a bounded support. But then its HF contribution into the asymptotics (2) is negligibly small or, more accurately, $|\hat{f}(\mathbf{p}\omega|\mathbf{x}_0)| < C\omega^{-n}$, if the order of smoothness $n$ of $f(\mathbf{x})$ is finite. This is a well-known fact from Fourier transform theory (Fedoryuk, 1987).

It follows that the HF asymptotics of $\hat{u}(\omega)$ in the smooth case is controlled by properties of the source function near the boundary of the support $\Omega_f$ and by properties of the boundary $\partial \Omega_f$ itself.

We will call $\partial\Omega_f$ the frontal surface of the seismic source and represent it as $t = t_r(\mathbf{g})$ in a suitable neighborhood of $\mathbf{x}_0 \in \partial\Omega_f$. In addition, we will demand from $t_r(\mathbf{g})$ smoothness or piecewise smoothness near $\mathbf{x}_0$.

In the future we will operate with the time-arriving function

$$t_a(\mathbf{g}) = t_0 - <\mathbf{g}, \boldsymbol{\gamma}_\Sigma/c> + t_r(\mathbf{g}) \qquad (4)$$

which determines the travel time of the seismic signal from the frontal point to the remote receiver. Isolines of $t_a(\mathbf{g})$ will be termed isochrons.

The behavior of the source function near the rupture front has not been fully formalized in the literature. It is assumed that the source function is bounded during the rupture-healing phase [Madariaga et al., 1998; Nielsen and Madariaga, 2003]. As to the active rupture phase, the classical models by Kostrov [1964] and Madariaga [1977, 1983] demonstrate singularities of the 'inverse square root' type, $1/\sqrt{\cdot}$. A rigorous justification of the singularity in the framework of elasticity theory can be found in [Herrero et al., 2006]. Combining the two cases, we shall consider the source function around $\mathbf{x}_0$ to be

$$f(\mathbf{g}, t) = (t - t_r(\mathbf{g}))_+^{\beta-1} V(\mathbf{g}, t), \qquad 0 < \beta \leq 1, \qquad (5)$$

where $a_+ = \max(a, 0)$. The physically interesting cases are $\beta = 1$ and $\beta = 1/2$. In the first of these, we have a bounded $f$, while the second involves a singularity of the inverse square root type.

The two statements to follow are corollaries from Erdelyi's Lemma and from the method of stationary phase (see Appendix A1).

*Statement 1.* Assume that, in a vicinity of $\mathbf{x}_0 = (\mathbf{g}_0, t_0) \in \partial\Omega_f$, the seismic source function $f(\mathbf{x})$ can be represented as (5) with smooth components $t_r(\mathbf{g})$ and $V(\mathbf{g}, t) \neq 0$. Assume further that $t_r(\mathbf{g})$ has an isolated, *stationary, regular point* $\mathbf{g}_s$, i.e., $\mathbf{g}_s$ is unique in a vicinity of $\mathbf{g}_0$, such that $\nabla t_a(\mathbf{g}_s) = 0$ and the full curvature of the surface $t_a(\mathbf{g})$ at $\mathbf{g}_s$ is nonzero: $K_s \neq 0$. Then the HF contribution of $\mathbf{x}_0$ in the source spectrum has order $O(\omega^{-(\beta+1)})$, and the spectral amplitude depends on radiation directivity via the stationary point. To be more specific,

$$\hat{u}(\omega \mid \mathbf{x_0}) \approx C_\beta \omega^{-(1+\beta)} 2\pi A |K_s|^{-1/2} V(\mathbf{g}_s, t_r(\mathbf{g}_s)\varepsilon + o(\omega^{-2}) \ , \ |\varepsilon|=1 \ , \ \omega \to \infty \ . \tag{6}$$

The desired quadratic decay of the spectrum occurs with $\beta=1$, i.e., when $f(\mathbf{x})$ is bounded near $\mathbf{g}_0$.

If there are no stationary points near $\mathbf{g}_0$, then its HF contribution in the spectrum is of order $o(\omega^{-2})$ for any $\beta > 0$.

The details of the proof of (6) can be found in Appendix A2.

***The supershear effect***. The existence of a regular stationary point $\mathbf{g}_s$ for $t_a(\mathbf{g})$ means that

$$\nabla t_r(\mathbf{g}_s | \mathbf{x}_0) = \boldsymbol{\gamma}_\Sigma / c \ . \tag{7}$$

Denote by $\mathbf{v}(\mathbf{g}) = (v^1(\mathbf{g}), v^2(\mathbf{g}))$ the front velocity at $\mathbf{g}$. Differentiating the equality $t = t_r(\mathbf{g}|\mathbf{x}_0)$ by t and taking into account the relations $|\boldsymbol{\gamma}_\Sigma| \leq 1$, we get a *necessary condition* for the existence of a stationary point:

$$1 = <\nabla t_r(\mathbf{g}_s|\mathbf{x}_0), \mathbf{v}(\mathbf{g}_s)> = <\boldsymbol{\gamma}_\Sigma / c, \mathbf{v}(\mathbf{g}_s)> \leq |\mathbf{v}(\mathbf{g}_s)|/c, \tag{8}$$

where $v_r(\mathbf{g}_s)$ is the full velocity value.

The above inequality means that the rupture at a stationary point must move faster than wave does: $v_r(\mathbf{g}_s) \geq c$. This supershear phenomenon is possible, although rarely observed for earthquakes [Madariaga et al., 2000]. Now it is the object of a fairly large number of studies, analytical, numerical, and experimental (see e.g., [Rosakis et al., 1999; Dunham&Archuleta,2004; Yang et al,2004; Bizzari et al,2010; Marty et al, 2019]).

The classical model of Kostrov [1964] has the seismic source function of the form $f(\mathbf{g},t) = A \cdot d/dt(v_r^2 t^2 - |\mathbf{g}|^2)_+^{1/2}$, hence the frontal surface is a cone: $(v_r t)^2 = |\mathbf{g}|^2$. The curvature of a conical surface is zero. It follows that, according to (6), the HP contributions due to all stationary points of a conical frontal surface into the $\omega^{-2}$ asymptotics are zero.

We now make the requirement for the existence of a stationary point at the frontal surface $t_r(\mathbf{g})$ less stringent, replacing stationarity with *conditional stationarity*. Suppose $t_r(\mathbf{g})$ loses smoothness at a line $l: g_2 = \psi(g_1)$. Assume that $t_r(\mathbf{g})$ can be represented by two smooth surfaces $t_r^\pm(\mathbf{g})$ having a common smooth edge above the smooth curve $l$. Consider a contraction of $t_a(\mathbf{g})$ to the line $l: S(g_1) = t_a(g_1, \psi(g_1))$. The point $\tilde{\mathbf{g}}_s = (\tilde{g}_{1s}, \psi(\tilde{g}_{1s}))$ will be referred to as a conditional stationary point at $t_r(\mathbf{g})$, if $S'(\tilde{g}_{1s}) = 0$. A conditional stationary point is considered to be regular, if

$$S'(\tilde{g}_{1s}) = 0, \qquad \tilde{K}_s := S''(\tilde{g}_{1s}) \neq 0. \tag{9}$$

***Statement 2.*** Assume that, in a vicinity of $\mathbf{x}_0 = (\mathbf{g}_0, t_0) \in \partial\Omega_f$, the seismic source function $f(\mathbf{x})$ has representation (5) with a smooth component $V(\mathbf{g},t) \neq 0$. and the piecewise smooth rupture front $t_r(\mathbf{g})$ indicated above. Assume further that there is a unique conditional regular stationary point $\tilde{\mathbf{g}}_s = (\tilde{g}_{1s}, \psi(\tilde{g}_{1s}))$ near $\mathbf{x}_0 = (\mathbf{g}_0, t_0)$. Then the HF contribution of $\mathbf{x}_0$ in the source spectrum has order $O(\omega^{-(\beta+3/2)})$, and the spectral amplitude depends on radiation directivity via the stationary point. More specifically,

$$\hat{u}(\omega | \mathbf{x}_0) \approx C_\beta \omega^{-3/2-\beta} \cdot A\sqrt{2\pi} \left|\tilde{K}_s\right|^{-1/2} V(\tilde{\mathbf{g}}_s, t_r(\tilde{\mathbf{g}}_s)) \varepsilon \Delta \quad |\varepsilon| = 1, \tag{10}$$

where

$$\Delta = [\partial/\partial g_2 t_r^+(\tilde{\mathbf{g}}_s) - \gamma_\Sigma^{(2)}/c]^{-1} - [\partial/\partial g_2 t_r^-(\tilde{\mathbf{g}}_s) - \gamma_\Sigma^{(2)}/c]^{-1}. \tag{11}$$

Under conditions $\nabla t_r^\pm(\mathbf{g}|\mathbf{x}_0) \neq \gamma_\Sigma/c$, the quantity $\Delta$ is bounded and non zero.

The details of the proof of (10) can be found in Appendix A3.

***Remarks.*** In the case under consideration,

- the $\omega^{-2}$ behavior of the spectrum occurs for a physically natural value of the parameter $\beta = 1/2$;

- *the admissible directions* at which the asymptotics (10) can occur are either absent or make a straight segment on the rupture plane.

Indeed, the conditional stationary point $\tilde{\mathbf{g}}_s$ is an extreme point of $t_a(\mathbf{g})$, provided the function is considered at the line $l: L(\mathbf{g}) = 0$. Applying the Lagrange method to the conditional extremum, we conclude that the point $\tilde{\mathbf{g}}_s$ along with the unknown constant $\lambda$ are given by the relation

$$\nabla t_r^+(\mathbf{g}) - \hat{\boldsymbol{\gamma}}_\Sigma / c = \lambda \nabla L(\mathbf{g}), \quad \mathbf{g} \in l. \tag{12}$$

It remains to notice the following. If we assume that equation (12) can be solved for the $\hat{\boldsymbol{\gamma}}_\Sigma$ vector, then it can also be solved for any vector

$$\hat{\boldsymbol{\gamma}}_\Sigma^\varepsilon = \hat{\boldsymbol{\gamma}}_\Sigma + \varepsilon \lambda c \nabla L(\mathbf{g}_s), \quad \left|\hat{\boldsymbol{\gamma}}_\Sigma^\varepsilon\right| \leq 1 \tag{13}$$

that involves a new constant, $\lambda^\varepsilon = \lambda - \varepsilon$.

4. **Fractal source functions**.

There is a different integral representation of the far-zone displacement that is formally identical with (1); in this representation the original function $f$ is treated as a local stress drop and $\Sigma$ as an area occupied by initial asperities [Kostrov and Das, 1988; Boatwright, 1988; Gusev, 1989]. In this case the physical restrictions on $f$ are rather hazy. Mikumo and Miyatake [1979] studied numerically the problem of random variation of stress drop on a fault. The fracture process was found to be quite chaotic, without any discernible rupture front. This circumstance has stimulated a stochastic or fractal approach to kinematic source models.

Gusev [2014] made a numerical study of the following model:

$$f = V(\mathbf{g})\phi(t - t_r(\mathbf{g})), \tag{14}$$

where $\phi(t) \geq 0$ is a smooth function on the semi-axis $t > 0$, $\phi(0) \neq 0 \cup \infty$ and $\phi(t) = 0$ outside of $[0, t_0]$; $V(\mathbf{g})$ is a time-independent local stress drop at $\mathbf{g}$ and $t_r(\mathbf{g})$ the time of this event; both $V(\mathbf{g})$ and $t_r(\mathbf{g})$ are continuous. The rupture front is almost flat and is unidirectional:

$$t_r(\mathbf{g}) = <\mathbf{g}, \boldsymbol{\gamma}_r> / v + \delta t_r(\mathbf{g}), \tag{15}$$

Here, $\boldsymbol{\gamma}_r$, $|\boldsymbol{\gamma}_r| = 1$ is the dominant direction of rupture, and $\delta t_r(\mathbf{g})$ is a small perturbation. The support of $f$ is

$$\Omega_f : t_r(\mathbf{g}) < t < t_r(\mathbf{g}) + t_0 \tag{16}$$

The function $f$ is smooth over $t$ for inner points of $\Omega_f$. Consequently, the main contribution into the asymptotics of $\hat{u}(\omega)$ is due to points of the frontal surface $\partial \Omega_f$. Moreover, it comes from the first boundary: $t = t_r(\mathbf{g})$, because here $f$ is discontinuous as a function in the entire space.

In the Gusev [2014] model, the components $t_r(\mathbf{g})$ and $V(\mathbf{g})$ are fractal, i.e., their order of smoothness is below 1. Such objects are more conveniently described in terms of random functions. In that case the desired object is immersed in a population of similar objects with suitable physical properties (in our case, with a given fractional smoothness).

We shall say that a random function $\xi(\mathbf{g})$ has a smoothness of order $0 < H_\xi < 1$, if

$$E|\xi(\mathbf{g}) - \xi(\tilde{\mathbf{g}})|^2 \sim C(\mathbf{g})|\mathbf{g} - \tilde{\mathbf{g}}|^{2H_\xi}, \qquad \mathbf{g} - \tilde{\mathbf{g}} \to 0, \qquad (17)$$

where $E$ denotes mathematical expectation and $C(\mathbf{g}) > 0$ is a smooth function; the exponent $H_\xi$ is commonly referred to as the Hurst parameter.

Considering the random models, we have to replace the individual spectrum $\hat{u}(\omega)$ by the root-mean-square spectrum:

$$r.m.s.\hat{u}(\omega) = (E|\hat{u}(\omega)|^2)^{1/2}. \qquad (18)$$

Given (14,15), one has

$$\hat{f}(\mathbf{p}\omega) = \hat{\phi}(\omega) \int e^{i(\delta t_r(\mathbf{g}) + \langle \mathbf{D}, \mathbf{g} \rangle)\omega} V(\mathbf{g}) d\mathbf{g}, \quad \mathbf{D} = \gamma_r / v - \gamma_\Sigma / c. \qquad (19)$$

Hence, assuming the random components $\delta t_r(\mathbf{g})$ and $V(\mathbf{g})$ to be statistically independent, we get the main spectral relation for the doubly stochastic model (14):

$$E|\hat{u}(\omega)|^2 = |A|^2 |\hat{\phi}(\omega)|^2 \iint m_V(\mathbf{g}_1, \mathbf{g}_2) \chi_r(\omega|\mathbf{g}_1, \mathbf{g}_2) e^{i\omega \langle \mathbf{D}, \mathbf{g}_1 - \mathbf{g}_2 \rangle} d\mathbf{g}_1 d\mathbf{g}_2. \qquad (20)$$

Here, $m_V(\mathbf{g}_1, \mathbf{g}_2) = EV(\mathbf{g}_1)V(\mathbf{g}_2)$ is the correlation function of the $V(\mathbf{g})$ field, while $\chi_r$ is the characteristic function of the increments $\delta t_r(\mathbf{g})$:

$$\chi_r(\omega|\mathbf{g}_1, \mathbf{g}_2) = E e^{i\omega \xi}, \qquad \xi = \delta t_r(\mathbf{g}_1) - \delta t_r(\mathbf{g}_2). \qquad (21)$$

The relation (20) is the key to understanding why the source spectrum can be weakly dependent of direction $\mathbf{D}$. Assume that the distribution of $\xi = \delta t_r(\mathbf{g}_1) - \delta t_r(\mathbf{g}_2)$ has a smooth rapidly decreasing density. In that case its characteristic function $\chi_r(\omega|\mathbf{g}_1,\mathbf{g}_2)$ rapidly decreases over frequency, when $\mathbf{g}_1 \neq \mathbf{g}_2$. For this reason the HF asymptotics of (20) is controlled by a small vicinity of the diagonal $\mathbf{g}_1 = \mathbf{g}_2$, where the integrand $\exp(i\omega<\mathbf{D},\mathbf{g}_1-\mathbf{g}_2>)$ is weakly dependent on $\mathbf{D}$ directivity. The idea can be implemented in a broad class of distributions of $\xi$. For the sake of simplicity in the subsequent argument, the distribution will be assumed Gaussian in what follows.

***Statement 3***. Suppose the following requirements are satisfied:

(a) the random components of (14,15)), $V(\mathbf{g})$ and $\delta t_r(\mathbf{g})$, are statistically independent;

(b) the stress drop field $V(\mathbf{g})$ is such as to make $m_V(\mathbf{g}_1,\mathbf{g}_2) > 0$, $m_V(\mathbf{g},\mathbf{g}) < C$, and $E(V(\mathbf{g}_1)-V(\mathbf{g}_2))^2 \leq C|\mathbf{g}_1-\mathbf{g}_2|^{2h}, h > 0$ be true;

(c) the increments of $\delta t_r(\mathbf{g})$ have a Gaussian distributions with zero mean and the variance

$$E(\delta t_r(\mathbf{g}_1)-\delta t_r(\mathbf{g}_2))^2 = |\mathbf{g}_1-\mathbf{g}_2|^{2H}\sigma^2(\mathbf{g}_1,\mathbf{g}_2), \tag{22}$$

where $\sigma(\mathbf{g}_1,\mathbf{g}_2)$ is a smooth function that is bounded away from zero and infinity.

(d) $\phi(t) = t_+^{\beta-1}\varphi(t)$, where $0 < \beta \leq 1$, while $\varphi(t) \geq 0$, is a smooth finite function, and $\varphi(0) > 0$.

In that case we shall have

$$r.m.s.\hat{u}(\omega) \approx \omega^{-(\beta+1/H)}K_\beta \int m_V(\mathbf{g},\mathbf{g})\sigma^{-2/H}(\mathbf{g},\mathbf{g})d\Sigma, \qquad \omega \to \infty, \tag{23}$$

i.e., the HF asymptotics of the source spectrum is a power law function and is independent of radiation directivity.

***Remarks:***

-According to (23), the HF spectrum incorporates correlations among the random components of the model along the diagonal only. That means that the HF asymptotics of the spectrum is due to incoherent radiation from points of the source. A. Gusev emphasized that this property is necessary.

-The Gusev model arises, when $\beta = 1$. In this case a slightly rough frontal surface, i.e., a surface with the Hurst parameter near 1, is sure to generate a near-omega-square behavior of the spectrum. The situation is radically different in the limiting case, when $H = 1$, because the frontal surface can then be smooth. Statement 1 shows that the spectral quadratic decay is then provided by regular stationary (and unstable, as a matter of fact) points at the frontal surface. With the help of a slight roughness ($H \approx 1$) we stabilize, broadly speaking, the field of 'stationary points', thus obtaining an omega-square behavior that persists for nearly all directions. Some additional constraints are required to remove radiation directivity.

-All the above also applies to another physically interesting case, when $\beta = 1/2$. An omega-square behavior for this case occurs with a moderately ragged front, namely, when the Hurst parameter is $H = 2/3$.

## 5. Discussion: isochrons and asymptotics

Now we will try to informally explain the HF asymptotics found above. In our models, the source function near the frontal surface had a singularity of the type $(t - t_r(\mathbf{g}))_+^{\beta-1}$ with $\beta = 1$ or $\beta = 1/2$. For this reason the Fourier transform taken over time gives a HF contribution of order $\omega^{-\beta}$ (Erdelyi's Lemma, A1). An extra factor is due to the Fourier transform taken over space. What is important here is the analytical structure of the function $t_a(\mathbf{g}) = t_0 - <\mathbf{g}, \mathbf{\gamma}_\Sigma / c> + t_r(\mathbf{g})$, i.e., of the time the signal takes to travel from a point of rupture to a distant receiver in the direction $\mathbf{\gamma}_\Sigma$. This function plays the part of phase in a Fourier integral of the form

$$I(\varphi) = \int e^{i\omega t_a(\mathbf{g})} \varphi(\mathbf{g}) d\mathbf{g}. \tag{24}$$

The part played by a smooth function $\varphi$ with bounded support in the asymptotics of the integral is insignificant, serving effectively to neglect the boundary of the integration domain. With this circumstance in mind, we now consider (24) with $\varphi = 1$ in a small vicinity $G_0$ around $\mathbf{g}_0$.

Denote by $A(t)$ the measure of points in $G_0$ where $t_a(\mathbf{g}) < t$. Then (24) can be rewritten as $I(1_{G_0}) = \int_{t_-}^{t_+} e^{i\omega t} dA(t)$. The density of the measure, $\dot{A}(t)$, can be interpreted as the intensity of isolines (isochrons) of the function $t_a(\mathbf{g})$ on the set $G_0$. Such an object is not new to seismologists (see [Spudich & Frazer, 1984]). Its importance stems from the fact that the analytical properties of the density $\dot{A}(t)$ are intimately related to the HF asymptotics of $I(1_{G_0})$.

Turning to examples, we begin by considering Statement 1. Here, $t_a(\mathbf{g})$ has a regular stationary point $\mathbf{g}_s$ in $G_0$. Choosing suitable coordinates, one can then assume that $\mathbf{g}_s = 0$ and $t_a(\mathbf{g}) = -(g_1^2 \pm g_2^2)$.

Consider a simpler, elliptical, case corresponding to the sign (+). In this case, the isochrons are like ellipses converging toward the stationary point (Fig.1a) and $A(t) \approx C - \pi(-t)_+$ in a small vicinity of $t = 0$. Consequently, $\dot{A}(t)$ has a discontinuity at time $t = 0$, and therefore the asymptotics of $I(1_{G_0})$ will be of order $\omega^{-1}$. As a result, the local contribution in the spectrum asymptotics will be of order $\omega^{-\beta} \cdot \omega^{-1}$.

The hyperbolic case, $t_a(\mathbf{g}) = -(g_1^2 - g_2^2)$, is more time-consuming. In this case the isochrones are schematically depicted in Fig.1b, $\dot{A}(t) = c \ln|t|, |t| \ll 1$, and our previous conclusions about the spectrum remain valid

In Statement 2, the space is divided into two parts by the line $l$. In each of these one can choose suitable coordinates such that the line $l$ is given by relation $g_2 = 0$, the stationary point on $l$ is $\tilde{\mathbf{g}}_s = (0,0)$, $t_a(\mathbf{g}) = g_1^2 + g_2$ for $g_2 \leq 0$ and $t_a(g_1, g_2) = -t_a(g_1, -g_2)$ for $g_2 \geq 0$. In this example, the isochrones (Fig1c) first traverse the line $l$. Their traces on the line $l$ surround, and converge to, $\tilde{\mathbf{g}}_s$; at the critical time t=0, the isochrons touch the line $l$, and then move away from it. It is easy to show that $A(t) \approx C - k(-t)_+^{3/2}$ for small t. According to Erdelyi's Lemma (Appendix A1), the Fourier transform of $\dot{A}(t)$ has HF asymptotics of the order $\omega^{-3/2}$. Therefore, the local contribution in the source spectrum asymptotics will be of order $\omega^{-\beta} \cdot \omega^{-3/2}$.

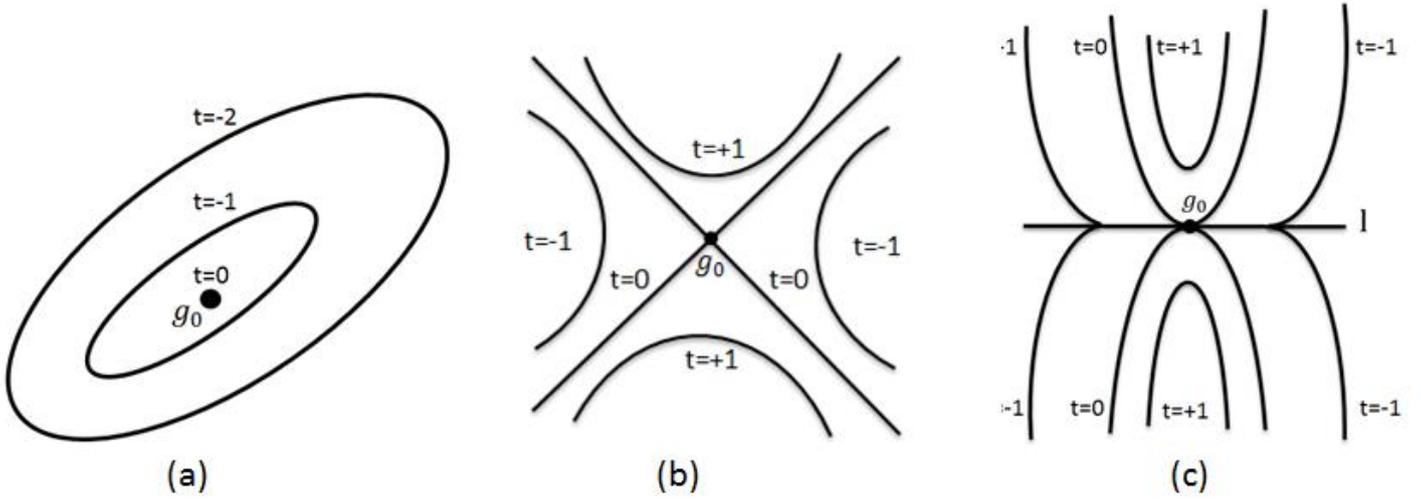

**Figure 1**. Isochrones of the arrival-time function $t_a(\mathbf{g})$ near a critical point $g_0$: (a),(b) are cases in which $g_0$ is a regular stationary point ; (c) is the case where $g_0$ is a conditional regular stationary point on the line $l$, where $t_a(\mathbf{g})$ loses smoothness

In Statement 3, the $t_a(\mathbf{g})$ field has a fractional smoothness of order $H<1$ at each spatial point, and is assumed to be random. When the density of isochrons is referred to in stochastic analysis, it has special name, viz., 'occupation density'. Under the assumptions we have made, the density $\dot{A}(t)$ is defined for almost all samples in the population [Xiao, 2013]. As well, the smoothness $\dot{A}(t)$ increases indefinitely, as the smoothness of the original field $t_a(\mathbf{g})$ decreases [Geman and Horowitz, 1978]. For this reason the part played by roughness of the frontal surface in the HF asymptotics of the source spectrum must decrease with increasing front raggedness (i.e. the Hurst parameter $H \to 0$). This is reflected in the final HF asymptotics $C\omega^{-\beta-1/H}$ of seismic source spectrum found in Statement 3.

A heuristic explanation of this result is based on the following: 1) HP asymptotics of the integral $I(1_{G_0}) = \int_{t_-}^{t_+} e^{i\omega t} dA(t)$ depends mainly on the end points $t_\pm$ 2) if an undisturbed linear front is propagating along the $g_1$-axis, then $dA(t) \approx Cdg_1$ near $t_+ = \max t$, and 3) due to random perturbation of the front with property (17) one has $dt \approx K(dg_1)^H$. As a result, $dA(t) \approx \tilde{C}(dt)^{1/H}$, i.e., $\dot{A}(t) \approx \tilde{C}(t_+ - t)_+^{-1+1/H}$ for $t$ close to $t_+$. Erdelyi's Lemma gives the desired order of the asymptotics ($\omega^{-1/H}$) for $I(1_{G_0})$. As above, the local contribution in the source spectrum asymptotics will be of order $\omega^{-\beta} \cdot \omega^{-1/H}$.

*Additional remarks*.

- The stochastic version of the source function has to deal with the root-mean-square source spectrum. Therefore, our conditions related to the loss of radiation directivity in the far zone are formulated in probabilistic terms that is convenient for modeling. The condition consists in the requirement that the distribution of increments of $t_a(\mathbf{g})$ at any pair of points in the space have a high order of smoothness. For example, the Rayleigh distribution with density $cx_+ \exp(-x^2/\sigma)$, which was used by Gusev [2014] for the random component of the rupture front, does not have that property It is not quite clear how to transfer the specified condition for the ensemble to an individual model of $t_a(\mathbf{g})$.

- The frequency range of the $\omega^{-2}$ behavior of the source spectrum is beyond the scope of the presented analysis. The issue must be dealt with numerically and was studied by A. Gusev for concrete models [Gusev, 2014].

## 6. Conclusions.

We have considered smooth and fractal local singularities in the seismic source function. Singularities of the first type enable, under definite conditions, reproduction of the far-zone omega-square behavior of the source spectrum. The source of such behavior may be the condition under which the velocity of the fracture front exceeds the P / S wave velocity. The asymptotic behavior of the spectrum for smooth models depends on radiation directivity, with the local source of power-law asymptotics being at best observable at receivers installed along an arc segment.

Fractality in source function models is a stabilizing factor. In that case one can reproduce both the quadratic decay of the spectrum and the loss of radiation directivity in the far zone. These properties can be reproduced (roughly speaking) during both the active rupture phase and during the healing phase. However, the respective requirements on the smoothness of the frontal surface are not the same: for the active rupture phase the smoothness is lower ($H = 2/3$ vs $H \approx 1$), which is quite natural physically. However, in both of these cases a high degree of non-smoothness ($H < 1/2$) in the frontal surface is incompatible with the omega-square behavior of the spectrum.

**Acknowledgments.** I am sincerely grateful to my colleagues A.Lander and A.Skorkina for careful reading of the work and useful comments.

# Appendix
**A1. Auxiliary statements [Fedoryuk, 1987].**

***Erdelyi's Lemma***. If $\varphi(t)$ is a smooth function with a bounded support, then for $\beta > 0$:

$$\int_0^\infty e^{i\omega t} t^{\beta-1} \varphi(t) dt = \omega^{-\beta} \varphi(0) \Gamma(\beta) e^{i\pi\beta/2} + \sum_1^N a_k \omega^{-(\beta+k)} + \delta_N(\omega), \omega \to \infty ,\qquad (A1)$$

where

$$\delta_N(\omega) < C\omega^{-[\beta+N]} \max \left| (d/dt)^{[\beta+N]} \varphi(t) \right|$$

and $[a]$ is the integer part of $a$.

***The method of stationary phase***. If smooth functions $f(\mathbf{x})$ and $S(\mathbf{x})$, $\mathbf{x} \in R^n$ are such that $f = 0$ outside of a bounded domain $\Omega \subset R^n$ and $S$ has a unique regular stationary point $\mathbf{x}_s$ in $\Omega$, i.e.,

$$\nabla S(\mathbf{x}_s) = 0 \text{ and } K(\mathbf{x}_s) = \det[\partial^2 / \partial x_i \partial x_j S(\mathbf{x}_s)] \neq 0 ,$$

then

$$\int_\Omega f(\mathbf{x}) e^{i\omega S(\mathbf{x})} d\mathbf{x} = a_s \omega^{-n/2} e^{i\omega S(\mathbf{x}_s)} (f(\mathbf{x}_s) + O(\omega^{-1})), \qquad \omega \to \infty \qquad (A2)$$

where

$$a_s = (2\pi)^{n/2} |K(\mathbf{x}_s)|^{-1/2} \exp(i\pi \operatorname{sgn} K(\mathbf{x}_s)/4) . \qquad (A3)$$

If $S$ has no stationary points in $\Omega$, $\nabla S(\mathbf{x}) \neq 0$, then the right-hand part of (A2) is negligibly small.

**A2. *Proof of Statement 1***. We assume that $f(\mathbf{g},t) = (t-t_r(\mathbf{g}))_+^{\beta-1} V(\mathbf{g},t)$ in a vicinity $O(\mathbf{x}_0)$ of $\mathbf{x}_0 \in \partial\Omega_f$ with smooth components $t_r(\mathbf{g}), V(\mathbf{g},t)$. The contribution of the vicinity of $\mathbf{x}_0$ into the asymptotics of $\hat{u}(\omega)$ is

$$\hat{u}(\omega|\mathbf{x}_0) = \int e^{i\omega(t+<\mathbf{g},\tilde{\mathbf{p}}>)} f(\mathbf{g},t) \varphi(\mathbf{g},t) d\mathbf{g} dt , \qquad (A4)$$

where $\tilde{\mathbf{p}} = -\boldsymbol{\gamma}_\Sigma / c$, $\varphi = \phi(\mathbf{g})\psi(t)$ is an auxiliary smooth finite function that is concentrated in the vicinity of $\mathbf{x}_0$. Making the time substitution $\tau = t - t_r(\mathbf{g})$, we shall have

$$\hat{u}(\omega|\mathbf{x_0}) = \int e^{i\omega(t_r(\mathbf{g})+<\mathbf{g},\tilde{\mathbf{p}}>)} J(\omega|\mathbf{g}) \tilde{\varphi}(\mathbf{g}) d\mathbf{g} , \qquad (A5)$$

where

$$J(\omega|\mathbf{g}) = \int_0^\infty e^{i\omega\tau} \tau^{\beta-1} \tilde{V}(\mathbf{g},\tau+t_r(\mathbf{g})) d\tau \qquad (A6)$$

and $\tilde{V}(\mathbf{g},t) = V(\mathbf{g},t)\psi(t)$. According to Erdelyi's Lemma (A1),

$$J(\omega|\mathbf{g}) = C_\beta \omega^{-\beta}[\tilde{V}(\mathbf{g},t_r(\mathbf{g})) + i\beta\omega^{-1}\partial/\partial t \tilde{V}(\mathbf{g},t_r(\mathbf{g}))] + \delta_2(\omega) \quad , \tag{A7}$$

$$\delta_2(\omega) < c\omega^{-[\beta+2]} \max_{O(x_0)} |(\partial/\partial t)^2 V(\mathbf{g},t)| . \tag{A8}$$

Combining (A5) and (A7), we have

$$\hat{u}(\omega|\mathbf{x_0}) \approx \int e^{i\omega(t_r(\mathbf{g})+<\mathbf{g},\tilde{\mathbf{p}}>)}[\tilde{V}(\mathbf{g},t_r(\mathbf{g})) + i\beta\omega^{-1}\partial/\partial t \tilde{V}(\mathbf{g},t_r(\mathbf{g}))]\tilde{\varphi}(\mathbf{g})d\mathbf{g} \cdot C_\beta \omega^{-\beta} + O(\omega^{-[2+\beta]}). \tag{A9}$$

Suppose the function $t_a(\mathbf{g}) = t_0 - <\mathbf{g},\gamma_\Sigma/c> + t_r(\mathbf{g})$ has a single stationary point $\mathbf{g}_s$ in $O(\mathbf{g}_0)$ and the full curvature of $t_a(\mathbf{g})$ (or, which amounts to the same thing, the curvature of $t_r(\mathbf{g})$) at the point $\mathbf{g}_s$ is different from zero, $K_s \neq 0$. In that case we use the method of stationary phase (A2) for the case $n=2$ to get

$$\hat{u}(\omega|\mathbf{x_0}) \approx C_\beta \omega^{-(1+\beta)} 2\pi |K_s|^{-1/2} V(\mathbf{g}_s, t_r(\mathbf{g}_s))\varepsilon + O(\omega^{-[2+\beta]}) \quad , |\varepsilon|=1 . \tag{A10}$$

Suppose the frontal surface $t_r(\mathbf{g})$ has no stationary points in a vicinity of $O(\mathbf{x_0})$. In that case, according to the method of stationary phase, the first term in (A9) is negligibly small as $\omega \to \infty$. Consequently, we have $\hat{u}(\omega|\mathbf{x_0}) = o(\omega^{-2})$ for any $\beta > 0$.

**A3**. *Proof of Statement 2*. We continue the analysis of (A9) on the assumption that $t_r(\mathbf{g})$ consists of two smooth parts $t_r^\pm(\mathbf{g})$, in a vicinity of $\mathbf{x_0} \in \partial\Omega_f$ that have a common edge above the smooth curve $l: g_2 = \psi(g_1)$. In that case (A9) can be written down as

$$\hat{u}(\omega|\mathbf{x_0}) \approx C_\beta \omega^{-\beta} e^{-i\omega t_0} (\int_{O^+(\mathbf{g}_0)} + \int_{O^-(\mathbf{g}_0)}) e^{i\omega t_a(\mathbf{g})} \phi(\mathbf{g}) d\mathbf{g} , \tag{A11}$$

where $t_a(\mathbf{g}) = t_0 - <\mathbf{g},\gamma_\Sigma/c> + t_r^\pm(\mathbf{g}|\mathbf{x_0})$, $\phi(\mathbf{g})$ is a smooth function, and $\nabla t_a(\mathbf{g}) \neq 0$ inside $O^\pm(\mathbf{g}_0)$. The asymptotic expression for integrals like (A11) can be found in the book [Fedoryuk, 1987; Ch.3, Theorem 4.1] and has the form

$$\int_{O^\pm(\mathbf{g}_0)} e^{i\omega t_a(\mathbf{g})} \phi(\mathbf{g}) d\mathbf{g} \approx \pm \omega^{-3/2} i\sqrt{2\pi} e^{i\omega t_a(\tilde{\mathbf{g}}_\gamma) + i\pi/4\cdot s} |\tilde{K}_s|^{-1/2} (\partial/\partial g_2 t_a^\pm(\tilde{\mathbf{g}}_s))^{-1} (\phi(\tilde{\mathbf{g}}_s) + O(\omega^{-1})). \tag{A12}$$

Substitution of (A12) in (A11) yields (10,11).

**A4**. *Proof of Statement 3*. We write (20) as $E|\hat{u}(\omega)|^2 = |A|^2 |\hat{\phi}(\omega)|^2 I_\omega$. Here, $\hat{\phi}(\omega)$ is the Fourier transform of $\phi(t) = t^{\beta-1}\varphi(t)$, and

$$I_\omega = \iint m_V(\mathbf{g}_1,\mathbf{g}_2)1(\mathbf{g}_1)1(\mathbf{g}_2)\chi_r(\omega|\mathbf{g}_1,\mathbf{g}_2)e^{i\omega<\mathbf{D},\mathbf{g}_1-\mathbf{g}_2>}d\mathbf{g}_1 d\mathbf{g}_2 \tag{A13}$$

where, $1(\mathbf{g})$ is an indicator of the source function support (which is assumed to be bounded). Erdelyi's Lemma (see **A1**) gives

$$\hat{\phi}(\omega) \approx \omega^{-\beta}\varphi(0)\Gamma(\beta)e^{i\pi\beta/2}, \omega \to \infty . \tag{A14}$$

It remains to find the asymptotics of (A13). Divide the integration domain into two parts, $\Omega_\omega \cup \Omega_\omega^c$, where

$$\Omega_\omega = \{(\mathbf{g}_1,\mathbf{g}_2):|\mathbf{g}_1-\mathbf{g}_2|<\omega^{-1/H}\ln\omega\}, \tag{A15}$$

The contributions of each of these subsets into (A13) will be denoted $I(\Omega_\omega)$ and $I(\Omega_\omega^c)$.

We now estimate $I(\Omega_\omega^c)$.

The characteristic function $\chi_r(\omega|\mathbf{g}_1,\mathbf{g}_2) = Ee^{i\omega\xi}$ of the Gaussian random variable $\xi = \delta t_r(\mathbf{g}_1) - \delta t_r(\mathbf{g}_2)$ with variance $\sigma_\xi^2$ is $exp(-\omega^2\sigma_\xi^2/2)$. The variance $\sigma_\xi^2$ was defined in (22). We have assumed $\sigma_\xi^2/|\mathbf{g}_1-\mathbf{g}_2|^{2H}$ to be bounded from below and $m_V(\mathbf{g}_1,\mathbf{g}_2)$ from above in the source region. Hence

$$\left|I(\Omega_\omega^c)\right| < C\iint_{\Omega_\omega^c} \exp(-\omega^2 k|\mathbf{g}_1-\mathbf{g}_2|^{2H})1(\mathbf{g}_1)1(\mathbf{g}_2)d\mathbf{g}_1 d\mathbf{g}_2 <$$

$$< C_1 \int_{r>\omega^{-1/H}\ln\omega} exp(-\omega^2 k r^{2H})dr^2 = C_1\int_{\ln\omega}^{\infty} e^{-kx^{2H}}dx^2 \cdot \omega^{-2/H} = \omega^{-2/H}o(1) . \tag{A16}$$

We are going to estimate $I(\Omega_\omega)$.

For this purpose, we decompose the integral into two terms, $I(Q|\Omega_\omega)$ and $I(1|\Omega_\omega)$, replacing the exponent with the sum $e^{i\omega<\mathbf{D},\mathbf{g}_1-\mathbf{g}_2>} = Q(\mathbf{g}_1-\mathbf{g}_2)+1$. We shall have

$$I(Q|\Omega_\omega) = \iint_{\Omega_\omega} m_V(\mathbf{g}_1,\mathbf{g}_2)1(\mathbf{g}_1)1(\mathbf{g}_2)\chi_r(\omega|\mathbf{g}_1,\mathbf{g}_2)Q(\mathbf{g}_1-\mathbf{g}_2)d\mathbf{g}_1 d\mathbf{g}_2 . \tag{A17}$$

Here, $m_V, \chi_r$ are bounded. The inequality $|e^{ix}-1|<|x|$ entails $|Q(\mathbf{g}_1-\mathbf{g}_2)|\leq \omega|\mathbf{g}_1-\mathbf{g}_2|$. Hence

$$|I(Q|\Omega_\omega)| < C\int_{\Omega_\omega} \omega|\mathbf{g}_1-\mathbf{g}_2|1(\mathbf{g}_1)1(\mathbf{g}_2)d\mathbf{g}_1 d\mathbf{g}_2 <$$

$$< C_1 \int_0^{\omega^{-1?H}\ln\omega} \omega \cdot r^2 dr = \omega^{-2/H}o(1) . \tag{A18}$$

It remains to find the asymptotics of

$$I(1|\Omega_\omega) = \iint_{\Omega_\omega} m_V(\mathbf{g}_1, \mathbf{g}_2) 1(\mathbf{g}_1) 1(\mathbf{g}_2) \exp(-\omega^2 |\mathbf{g}_1 - \mathbf{g}_2|^{2H} \sigma(\mathbf{g}_1, \mathbf{g}_2)) d\mathbf{g}_1 d\mathbf{g}_2 \ . \tag{A19}$$

Since $\sigma(\mathbf{g}_1, \mathbf{g}_2)$ is smooth on the set $\Omega_\omega$, we have the following estimates:

$$0 < \sigma_-(\mathbf{g}_1, \mathbf{g}_1) < \sigma(\mathbf{g}_1, \mathbf{g}_2) < \sigma_+(\mathbf{g}_1, \mathbf{g}_1), \ \sigma_\pm(\mathbf{g}, \mathbf{g}) = \sigma(\mathbf{g}, \mathbf{g}) \pm \rho\varepsilon, \tag{A20}$$

where $\varepsilon = \omega^{-1/H} \ln \omega$. Replacing $\sigma$ with $\sigma_\pm$, we shall have two-sided estimates $I_\pm(1|\Omega_\omega)$ of $I(1|\Omega_\omega)$, because $m_V(\mathbf{g}_1, \mathbf{g}_2) > 0$.

Now, we make the change of variables $\mathbf{v} = \mathbf{g}_1, \mathbf{u} = \mathbf{g}_1 - \mathbf{g}_2$, and represent these estimates as follows: $I_\pm(1|\Omega_\omega) = I_\pm + R_\pm$.

Here, $I_\pm$ estimates the contribution due to the diagonal:

$$I_\pm = \iint_{|\mathbf{u}| < \varepsilon} m_V(\mathbf{v}, \mathbf{v}) 1(\mathbf{v}) \exp(-\omega^2 |\mathbf{u}|^{2H} \sigma_\pm^2(\mathbf{v}, \mathbf{v})/2) d\mathbf{v} d\mathbf{u} =$$
$$= \int m_V(\mathbf{v}, \mathbf{v}) \sigma_\pm^{-2/H}(\mathbf{v}, \mathbf{v}) 1(\mathbf{v}) A_\pm(\mathbf{v}) d\mathbf{v} \omega^{-2/H} \ , \tag{A21}$$

Where

$$A_\pm(\mathbf{v}) = \int_0^{\ln \omega \cdot \sigma_\pm^{1/H}(\mathbf{v},\mathbf{v})} x \exp(-x^2/2) dx = (\pi/2)^{1/2} + o(1), \quad \omega \to \infty. \tag{A22}$$

For our estimate of $R_\pm$ we make use of the Cauchy inequality:

$$|m_V(\mathbf{v}, \mathbf{v} - \mathbf{u}) - m_V(\mathbf{v}, \mathbf{v})|^2 \le EV^2(\mathbf{v}) E(V(\mathbf{v} - \mathbf{u}) - V(\mathbf{v}))^2 \le K|\mathbf{u}|^{2h}.$$

Consequently, similarly to the case (A21), we shall have

$$R_\pm < K \iint_{|\mathbf{u}|<\varepsilon} |\mathbf{u}|^h 1(\mathbf{v}) \exp(-\omega^2 |\mathbf{u}|^{2H} \sigma_\pm^2(\mathbf{v}, \mathbf{v})/2) d\mathbf{v} d\mathbf{u} < \varepsilon^h \int \sigma_\pm^{-2/H}(\mathbf{v}, \mathbf{v}) 1(\mathbf{v}) A_\pm(\mathbf{v}) d\mathbf{v} \omega^{-2/H}. \tag{A24}$$

Combining (A16, A18, A21, A22, and A24), we get the desired asymptotic expression:

$$I = \sqrt{\pi/2} \int m_V(\mathbf{v}, \mathbf{v}) \sigma^{-2/H}(\mathbf{v}, \mathbf{v}) d\mathbf{v} \cdot \omega^{-2/H}(1 + o(1)). \tag{A25}$$